\documentclass[lettersize,journal]{IEEEtran}
\usepackage{amsmath,amsfonts}
\usepackage{algorithmic}
\usepackage{array}
\usepackage[caption=false,font=normalsize,labelfont=sf,textfont=sf]{subfig}
\usepackage{textcomp}
\usepackage{stfloats}
\usepackage{url}
\usepackage{booktabs}
\usepackage{verbatim}
\usepackage{graphicx}
\def\BibTeX{{\rm B\kern-.05em{\sc i\kern-.025em b}\kern-.08em
    T\kern-.1667em\lower.7ex\hbox{E}\kern-.125emX}}
\usepackage{balance}
\begin{document}
\title{Online-Optimized Gated Radial Basis Function Neural Network-Based Adaptive Control}
\author{Mingcong Li
\thanks{The author is with the School of Automation, Beijing Institute of Technology, Beijing 100081, China (e-mail: limingcong0627@163.com).}
}

\maketitle

\begin{abstract}
Real-time adaptive control of nonlinear systems with unknown dynamics and time-varying disturbances demands precise modeling and robust parameter adaptation. While existing neural network-based strategies struggle with computational inefficiency or inadequate temporal dependencies, this study proposes a hybrid control framework integrating a Temporal-Gated Radial Basis Function (TGRBF) network with a nonlinear robust controller. The TGRBF synergizes radial basis function neural networks (RBFNNs) and gated recurrent units (GRUs) through dynamic gating, enabling efficient offline system identification and online temporal modeling with minimal parameter overhead (14.5\% increase vs. RBFNNs). During control execution, an event-triggered optimization mechanism activates momentum-explicit gradient descent to refine network parameters, leveraging historical data to suppress overfitting while maintaining real-time feasibility. Concurrently, the nonlinear controller adaptively tunes its gains via Jacobian-driven rules derived from the TGRBF model, ensuring rapid error convergence and disturbance rejection. Lyapunov-based analysis rigorously guarantees uniform ultimate boundedness of both tracking errors and adaptive parameters. Simulations on a nonlinear benchmark system demonstrate the framework's superiority: compared to PID and fixed-gain robust controllers, the proposed method reduces settling time by 14.2\%, limits overshoot to 10\%, and achieves 48.4\% lower integral time-weighted absolute error under dynamic disturbances. By unifying data-driven adaptability with stability-guaranteed control, this work advances real-time performance in partially observable, time-varying industrial systems .
\end{abstract}


\section{Introduction}
\IEEEPARstart{C}{ontrol} of nonlinear systems with unknown dynamic characteristics remains a critical challenge in industrial automation and robotics, particularly for applications that require real-time adaptability under time-varying disturbances and unmodelled dynamics~\cite{b1,b2,b3,b4,b5}. Traditional model-based approaches that rely on idealised assumptions frequently fail to capture the inherent nonlinearities and dynamic uncertainties present in real-world systems~\cite{b18}. Although deep-learning-driven adaptive strategies~\cite{b7,b8,b11} have demonstrated promise in addressing unmodelled dynamics, several limitations remain. Deep neural networks (DNNs) lack explicit mechanisms for modelling temporal state dependencies, which leads to sub-optimal performance in systems that exhibit hysteresis or delayed responses~\cite{b12,b16,b19}. Recurrent neural networks (RNNs) and long short-term memory (LSTM) networks possess temporal modelling capabilities, but their computational complexity and parameter redundancy hinder real-time deployment~\cite{b6,b9,b15}. Moreover, offline training of neural networks requires comprehensive datasets that span the entire operating envelope—a requirement that is often impractical in environments with limited sensor coverage or frequent dynamic variations~\cite{b10,b17}. Performance degradation is further aggravated by discrepancies between simulated training data and actual system dynamics~\cite{b11}. Current methodologies therefore emphasise offline model fitting and neglect the need for real-time parameter adaptation to cope with time-varying disturbances or system degradation~\cite{b13,b20}.

In the literature, \cite{b27} developed an adaptive state observer using feedforward-structured auto-regressive radial basis function (ARX-RBF) neural networks to estimate unknown states in a class of time-varying, delayed nonlinear systems. Peng et al.~\cite{b28} proposed a multi-input multi-output (MIMO) ARX-RBF neural network for nonlinear system modelling. This feedforward architecture demonstrated significant performance improvements over linear models, enabling enhanced control of nonlinear systems. However, its capacity for capturing long-term dependencies is limited, as its temporal modelling originates primarily from an explicit delay structure. Although recent advancements in knowledge-embedded deep neural networks have improved generalisation in model predictive control through the integration of domain-specific prior knowledge~\cite{b4}, their adaptability to dynamic environments remains constrained by a reliance on high-fidelity models~\cite{b4, b22}.

Offline-trained deep neural networks, such as those used for model predictive controller compensation~\cite{b3} or robotic arm friction modelling~\cite{b16}, exhibit limited generalisation under dynamic operating conditions due to their reliance on static datasets~\cite{b10,b17}. Even when augmented with online parameter updates~\cite{b13,b15}, the computational overhead of high-dimensional networks (e.g., LSTMs~\cite{b6,b8} and other dynamic neural architectures~\cite{b19}) restricts their feasibility in latency-sensitive applications. For instance, while satellite attitude control systems~\cite{b9} require rapid online adjustments of their RBFNN hidden layers, conventional RBFNNs lack mechanisms to balance approximation accuracy with real-time computational demands~\cite{b21}. Parameter redundancy exacerbates these issues, as seen in the semi-linearised DNNs for robotic systems~\cite{b16} and multi-layer neural controllers~\cite{b25}, where an excessive number of weights hinders deployment on embedded platforms. Notably, while an efficient deep-learning-based representation for model predictive control laws was proposed in~\cite{b23}, its adaptability in lifelong learning scenarios requires further investigation~\cite{b24}.

Integrating neural approximations into control frameworks with stability guarantees introduces substantial complexity. Although DNN-enhanced sliding mode control (SMC)~\cite{b22} and integral backstepping strategies~\cite{b20} improve robustness against uncertainties, they often require sophisticated chattering-suppression mechanisms or suffer from over-parameterisation. Lyapunov-stable LSTM controllers~\cite{b6} and physics-informed neural networks~\cite{b10} ensure stability through constrained training, but this necessitates offline pre-tuning, which in turn limits their adaptability to unforeseen disturbances. Similarly, the adaptive neural control method proposed in~\cite{b26}, while capable of handling unknown dynamics, relies on strong assumptions to guarantee stability. Hybrid model-predictive-control--DNN frameworks~\cite{b4,b3} excel at nonlinear compensation but depend on high-fidelity models for trajectory planning, which are unavailable in environments with completely unknown dynamics~\cite{b22}. Even advanced approaches, such as self-organising recurrent RBF networks~\cite{b1} and performance-guaranteed neural controllers~\cite{b25}, face a trade-off between adaptability and computational feasibility.

To address these challenges, we propose a Temporal-Gated Radial Basis Function (TGRBF) network that synergizes the rapid convergence of RBFNNs with the temporal memory capabilities of gated recurrent units (GRUs), incorporating a novel gradient descent method from \cite{b14} to optimize online learning efficiency. This architecture forms the foundation for a novel adaptive control strategy integrated with a high-performance controller. The principal innovations of this study encompass:

\begin{enumerate}
    \item Design of a lightweight hybrid network architecture (TGRBF) for system modeling: This network embeds GRU-based gating mechanisms within RBFNNs, achieving long-term temporal modeling with merely 14.5\% additional parameters compared to conventional RBFNNs. This architecture resolves the computational constraints of LSTM-based controllers  while surpassing the temporal modeling capacity of existing RBF networks .
    \item Development of an event-triggered online optimization framework: We implement momentum-explicit gradient descent for dynamic parameter adaptation of both TGRBF and controller parameters during real-time operation, significantly enhancing control precision and robustness.
    \item Formal stability guarantees through Lyapunov analysis: We establish rigorous proofs of bounded tracking errors and network parameter convergence, with comprehensive simulation studies validating the superior performance of the proposed control strategy.

\end{enumerate}

 This integrated approach addresses the critical gap between model-based control theory and data-driven adaptation, particularly for systems operating under time-varying dynamics and partial observability constraints. The theoretical framework ensures stability while maintaining computational efficiency suitable for real-time embedded implementations. 

The remainder of this paper is organized as follows. Section~II formulates the problem. Section~III introduces the fundamental concept of the TGRBF network and the proposed adaptive control strategy. Section~IV analyzes the stability of the TGRBF-NC. Section~V presents simulation results that demonstrate its effectiveness, and Section~VI draws the final conclusions.

\section{PROBLEM FORMULATION}

To ensure the general applicability of the proposed method, we consider a class of discrete-time nonlinear systems with unknown dynamics, whose input-output relationship is given by: 
\begin{equation}
y(t) = g(\xi(t-1)) + d(t)
\end{equation}

where the state vector \(\xi(t-1) = [y(t-1), \ldots, y(t-n_y), u(t-1-t_d), \ldots, u(t-n_u-t_d)]^T \in \mathbb{R}^{n_y+n_u}\) includes past outputs and delayed control inputs. The term \(u(t) \in \mathbb{R}\) is the control input, \(y(t) \in \mathbb{R}\) is the measured output, and \(d(t)\) is a time-varying bounded disturbance. The function \(g(\cdot) : \mathbb{R}^{n_y + n_u} \rightarrow \mathbb{R}\) is an unknown smooth nonlinear mapping.

\textbf{Assumption~1 (Local Lipschitz Continuity):} For any \(\xi_1,\,\xi_2 \in \mathcal{X}\;(\mathcal{X}\subset\mathbb{R}^{n_y+n_u})\), there exists a positive constant \(L_g\) such that
\begin{equation}
    \| g(\xi_1) - g(\xi_2) \| \leq L_g \| \xi_1 - \xi_2 \|
\end{equation}

\textbf{Assumption~2 (Bounded Disturbance Energy):} The disturbance \(d(t)\) satisfies \(\sum_{t=0}^{T} \lVert d(t) \rVert^2 \leq \bar{D}\), where \(\bar{D}\) is a constant.

\section{Design of the Temporal-Gated Radial Basis Function Network and Adaptive Control}

The following subsections detail the design of the TGRBF-NC method.

\subsection{TGRBF Network}

The TGRBF network is a hybrid architecture that integrates the rapid local approximation capability of a Radial Basis Function Network (RBFNN) with the temporal sequence modelling capability of a Gated Recurrent Unit (GRU) via a dynamic gating mechanism. This section details its design principles, mathematical implementation, and key innovations.

The RBF network is a three-layer feedforward neural network comprising an input layer, a hidden layer, and a linear output layer. The hidden layer employs radial basis functions as activation functions, enabling the network to approximate complex functions efficiently by mapping the input space to a higher-dimensional one. Owing to its simple structure, the network's complexity does not increase significantly with multi-variable inputs. Furthermore, its radial symmetry and the existence of arbitrary derivatives enhance its structural stability. For an input sample \(X = [x_1, x_2, \ldots]\), the Gaussian kernel function is expressed as: 
\begin{equation}\varphi_i(X,C_i)=\exp\left(-\frac{1}{2{b_i}^2}\|X-C_i\|^2\right)\end{equation}

where \(C_i\) and \(b_i\) represent the centre vector and width of the $i$-th neuron, respectively. These parameters control the local sensitivity of the basis function.

The output of the hidden layer is then linearly weighted to produce the final prediction: 
\begin{equation}
    y_{\text{rbf}}(k) = \sum_{i=1}^{m} w_i \varphi_i(X, C_i)
\end{equation}

where \(w_i\) is the weight connecting the $i$-th hidden node to the output layer, and \(m\) is the number of hidden nodes.

The RBF network, characterised by its "local activation" property, can efficiently approximate static nonlinear functions. However, it cannot model dynamic temporal dependencies. This branch serves as the "rapid-response module" of the TGRBF network, responsible for the instantaneous nonlinear mapping of the current input \(x(t)\).

The gating mechanism integrates historical data with the current input through update and reset gates, enabling the network to model temporal dependencies.

The update gate \(z_t\) is defined as:
\begin{equation}
    z_t = \sigma(W_z[x_t, h_{t-1}] + b_z)
\end{equation}

where \(\sigma(\cdot)\) is the sigmoid activation function, \(W_z\) and \(b_z\) are the corresponding learnable weight matrix and bias vector, and \(h_t \in \mathbb{R}^p\) is the hidden state. 

Similarly, the reset gate \(r_t\) is defined as:
\begin{equation}
    r_t = \sigma(W_r[x_t, h_{t-1}] + b_r)
\end{equation}

A candidate hidden state \(h_t'\) is then computed using the reset gate:
\begin{equation}
h_t' = \tanh(W_h[x_t, r_t \odot h_{t-1}] + b_h)
\end{equation}

where \(\tanh(\cdot)\) is the hyperbolic tangent activation function, \(W_h\) and \(b_h\) are its learnable weight matrix and bias vector, and \(\odot\) denotes the element-wise product. The final hidden state is updated as follows:
\begin{equation}
    h_t = (1 - z_t) \odot h_{t-1} + z_t \odot h_t'
\end{equation}

However, the nonlinear activation functions in the GRU complicate subsequent parameter updates; therefore, a linearised version (LGRU) is employed.

The LGRU is implemented as follows:
\begin{equation}
    r_t = \text{clamp}(r_t, 0, 1)
\end{equation}
\begin{equation}
    z_t = \text{clamp}(z_t, 0, 1)
\end{equation}
\begin{equation}
    n_t = n_t + r_t \odot h_{t-1}
\end{equation}
\begin{equation}
   h_t = (1 - z_t) \odot h_{t-1} + z_t \odot n_t
\end{equation}

To effectively combine the RBF and GRU branches, a dynamic gating mechanism is used to fuse their respective outputs. The predicted values from both branches are first concatenated into a single feature vector. A dynamic gating weight then adaptively balances the contributions of the two branches.

The gating weight \(g(t) \in [0,1]\) is generated from the concatenated vector \(\zeta_t = [x_t; h_{t-1}]\), which combines the current input \(x_t\) and the previous GRU hidden state \(h_{t-1}\), through a fully connected layer:
\begin{equation}
    g(t) = \sigma \left( w_g^T \zeta_t + b_g \right)
\end{equation}

where \(w_g\) is the weight vector and \(b_g\) is the bias. To prevent either branch from becoming dormant during training, the initial value of \(b_g\) is set to 0.5.

Finally, the overall TGRBF output is a weighted sum of the two branches, based on the dynamic gate \(g(t)\):
\begin{equation}
    y_{\text{tgrbf}}(t) = g(t) \cdot y_{\text{rbf}}(t) + (1 - g(t)) \cdot y_{\text{gru}}(t)
\end{equation}

\begin{figure}
    \centering
    \includegraphics[width=0.75\linewidth]{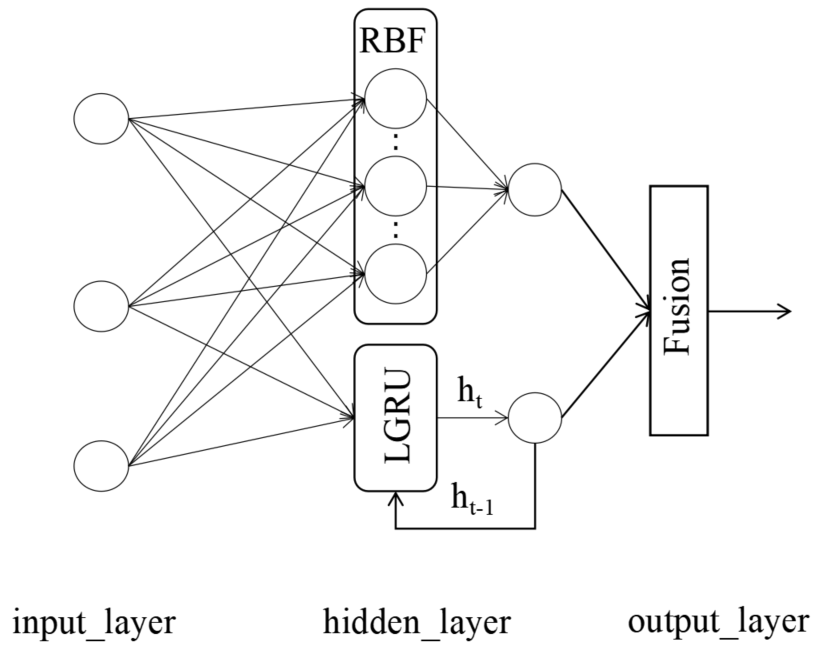}
    \caption{The architecture of the TGRBF network.}
    \label{fig1}
\end{figure}

The architecture of the TGRBF network is shown as Fig \ref{fig1} . 

 \subsection{Offline Identification Strategy}

 1) \textit{Data Acquisition and Pre-processing}: Collecting sufficient data from certain physical systems can be challenging due to operational constraints. Therefore, during the offline fitting stage, we first establish an idealised simulation model of the controlled object: 

 \begin{equation}
     y_{\text{sim}}(t) = g_{\text{nom}}\left(\xi(t-1)\right) + w(t)
 \end{equation}
Where, \( g_{\text{nom}} \) is the nominal model and \( w(t) \) is white noise that accounts for sensor noise and unmodelled dynamics.

By applying a continuous sequence of control inputs to this idealised model, datasets for various operating conditions can be generated. If the number of samples is sufficiently large and the system states comprehensively cover the actual operating range, the trained model can generalise to all required prediction tasks.

The simulated data are organised into a time-ordered dataset. To improve training outcomes, we employ the Teacher-Forcing strategy~\cite{b29}. Teacher-Forcing is a common training technique for sequence prediction tasks. Its core concept is to use the ground-truth values from the training dataset as input for the current time step, rather than the model's own output from the previous step. This approach improves training efficiency and stability by preventing the accumulation of prediction errors, thereby accelerating model convergence.

In this work, the TGRBF network is trained with an input vector \( \mathbf{x}_k = [u_k, y_{k-1}, y_k] \), which contains the control input \(u_k\), the previous system state \(y_{k-1}\), and the current system state \(y_k\). The network's output is the predicted current state, \(y_k\).

When the trained model is deployed for real-time control and testing, the current system state \(y_k\) is not available as an input. The input vector is therefore modified to \( \mathbf{x}_k = [u_k, y_{k-1}, y_{k-1}] \), where the previous state is used as a substitute for the current one. The output remains the predicted current state.

This offline training strategy accelerates training and reduces the demand for training samples while simultaneously improving the model's fitting accuracy. The benefits of this approach are particularly pronounced for multi-input, multi-output (MIMO) systems.

\subsection{Online Optimisation Strategy}
Since the offline fitting relies on an idealised simulation model, the trained network may still exhibit discrepancies when compared to the real system due to parameter inaccuracies, unmodelled dynamics, or external disturbances. It is therefore necessary to optimise the network continuously online. 

To meet the demands of real-time control, both in terms of computational speed and stability, we have designed an event-based explicit-step-size optimisation method to update the TGRBF network parameters. This method is based on the explicit gradient descent algorithm proposed in~\cite{b14} and incorporates historical data via a momentum term.

1) \textit{Event-Triggered Mechanism}: At each time step \(t\), the network's output error is defined as \( e_t = y_t^{\text{real}} - y_t^{\text{pred}} \). An update is triggered if \( \|e_t\| > \delta \), where \(\delta\) is a pre-defined threshold. When triggered, \(s\) data points, \(\left\{ (x_k, y_k) \right\}_{k=1}^s\), are sampled from an experience buffer, \(\mathcal{D}\), to update the network. The experience buffer has a capacity of 1000 samples and is populated with real-time system data, ensuring that the sampled data are independently and identically distributed. The buffer is updated using a first-in, first-out (FIFO) policy, with a preference for retaining high-error samples.

2) \textit{Network Parameter Update Algorithm}: The following loss function is defined:
\begin{equation}
    J(W) = \frac{1}{2s} \sum_{k=1}^{s} \left\| y_k - \hat{y}_k(W) \right\|^2
\end{equation}

Where,\(W = \left[ w_i, c_i, W_z, W_r, W_h, W_g \right]^T\) is the vector of parameters to be optimised .\( \hat{y}_k(W)\) is the predicted value of the network .

The explicit-step-size gradient descent method with a momentum term is adopted, with the parameter update rule given by: 
\begin{equation}
    W_{t+1} = W_t - \eta_t \nabla_W J(W_t) + \alpha (W_t - W_{t-1})
\end{equation}

The step size \(\eta_t\) is calculated as: 
\begin{equation}
    \eta_t = \frac{\mathbf{v}_t^T \mathbf{F}(\mathbf{W}_t)}{\mathbf{v}_t^T \mathbf{v}_t}, \quad \mathbf{v}_t = \nabla_{\mathbf{W}} \mathbf{F}(\mathbf{W}_t) \nabla_{\mathbf{W}} \mathbf{F}(\mathbf{W}_t)^T \mathbf{F}(\mathbf{W}_t)
\end{equation}

Where, \(\mathbf{F}(\mathbf{W}_t) = \left[ y_1 - \hat{y}_1(\mathbf{W}_t), \ldots, y_s - \hat{y}_s(\mathbf{W}_t) \right]^T\) is the residual vector, \(\nabla_{\mathbf{W}} \mathbf{F}(\mathbf{W}_t)\) is the Jacobian matrix .According to this update rule, we can get the specific parameter update formula of the designed TGRBF network: 
\begin{equation}
\begin{split}
    w_i^{(t+1)} = &w_i^{(t)} - \eta_t \cdot \frac{1}{s} \sum_{k=1}^{s} \left( \phi(x_k, c_i^{(t)}) \cdot g^{(t)}(x_k) \cdot e_k \right) \\&+ \alpha \left( w_i^{(t)} - w_i^{(t-1)} \right)
\end{split}
\end{equation}
\begin{equation}
\begin{split}
c_i^{(t+1)} =& c_i^{(t)} - \eta_t \cdot \frac{1}{s} \sum_{k=1}^{s} \left( w_i^{(t)} \cdot \frac{\partial \phi(x_k, c_i^{(t)})}{\partial c_i}  \cdot g^{(t)}(x_k) \cdot e_k \right)\\&+ \alpha \left( c_i^{(t)} - c_i^{(t-1)} \right)
\end{split}
\end{equation}
\begin{equation}
\begin{split}
    W_{z/r/h}^{(t+1)} = &W_{z/r/h}^{(t)} - \eta_t \cdot \frac{1}{s} \sum_{k=1}^{s} \left( \frac{\partial h_{\text{gru}}^{(t)}(x_k)}{\partial W_{z/r/h}} \cdot (1 - g^{(t)}(x_k)) \cdot e_k \right) \\&+ \alpha \left( W_{z/r/h}^{(t)} - W_{z/r/h}^{(t-1)} \right)
\end{split}
\end{equation}

\begin{figure}
    \centering
    \includegraphics[width=1\linewidth]{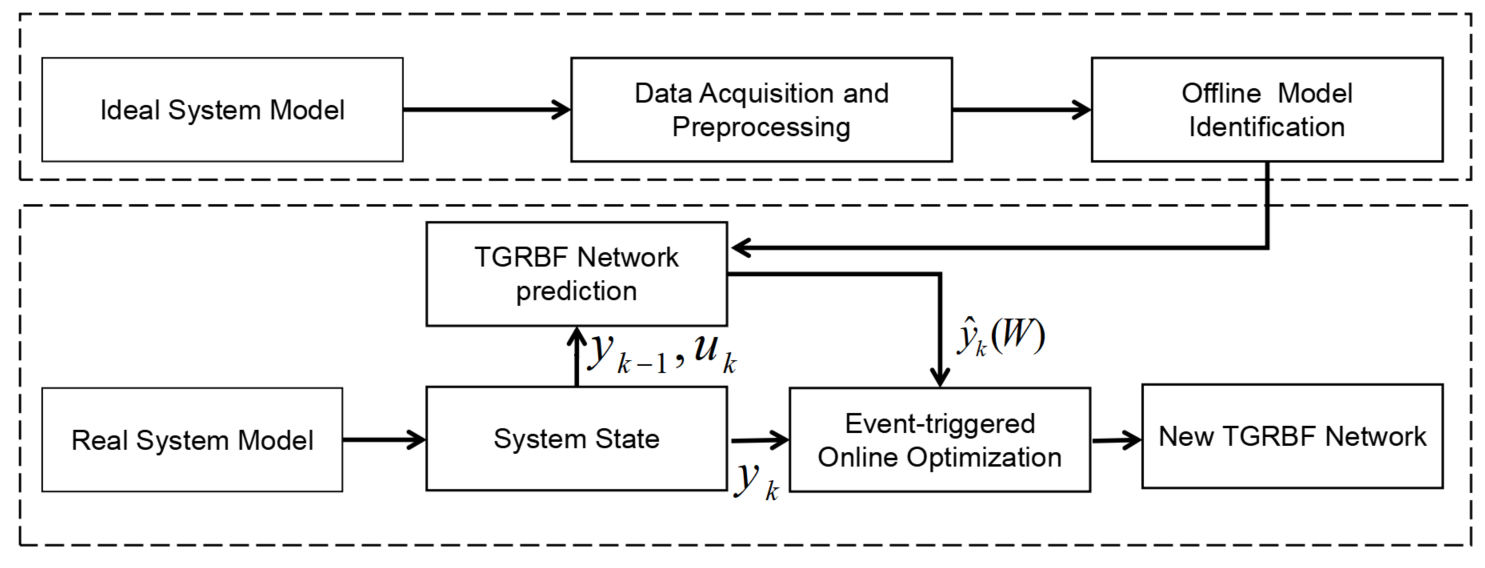}
    \caption{Flow of the TGRBF method}
    \label{fig2}
\end{figure}

The explicit step-size calculation enhances the stability of the parameter updates, while the momentum term improves the rate of gradient descent. The inclusion of momentum also helps to prevent overfitting to outlier data during online updates. Finally, the event-triggered strategy reduces the frequency of parameter updates, further decreasing the computational load. The overall fitting procedure is illustrated in Fig.~\ref{fig2}.

\subsection{Adaptive control method  }

After the model of the controlled system is obtained by fitting the designed tgrbf network, we further design the controller of the system. 

Considering the time-varying reference signal \(r (t) \in R\) that the controlled system needs to track, the tracking error is defined: 
\begin{equation}
    e(t)=r(t)-y(t)
\end{equation}

The controller we use is a nonlinear controller as :
\begin{equation}
    u=-k_1e-k_2sig^\alpha(e)
    \label{13}
\end{equation}

Here, \( k_1 \) and \( k_2 \) are the controller gains, and \( \alpha \in (0,1) \) is the power coefficient. The term \( -k_2 e \) is used for rapid convergence when the tracking error is large, while \( -k_2 \text{sig}(e) \) is employed to maintain system robustness and reduce high-frequency chattering.

Compared with traditional PID controllers, this controller requires fewer tuning parameters, exhibits superior control performance, and possesses finite-time stability.

We can utilize the fitted model network obtained in the previous section to optimize the gain parameters of this controller during the control process. For the gain parameters \( k_1 \) and \( k_2 \) of this controller, we employ the following adaptive update methods for optimization:
 \begin{equation}
     J_c = \frac{1}{2}[r(t) - y(t)]^2
 \end{equation}
 \begin{equation}
     k_i(t) = k_i(t-1) - \eta_i \frac{\partial J_c}{\partial k_i(t-1)}
 \end{equation}
 \begin{equation}
     \frac{\partial J_c}{\partial k_i(t-1)} = -e(t) \frac{\partial y}{\partial u(t-1)} \frac{\partial u(t-1)}{\partial k_i(t-1)}
 \end{equation}
 \begin{equation}
     \frac{\partial y(t)}{\partial u(t)} \approx \frac{\partial y_m(t)}{\partial u(t)}
 \end{equation}
 \begin{equation}
     k_1(t) = k_1(t-1) + \eta_2 e^2(t) \frac{\partial y_m}{\partial u}
     \label{n2}
 \end{equation}
 \begin{equation}
     k_2(t) = k_2(t-1) + \eta_2 e(t) \text{sig}^\alpha(t) \frac{\partial y_m}{\partial u}
     \label{n3}
 \end{equation}
 
 Computes Jacobian matrix from neural network model. Improved Hebb learning rules are used to guide the parameter tuning of nonlinear controller.
 
\begin{figure}
    \centering
    \includegraphics[width=1\linewidth]{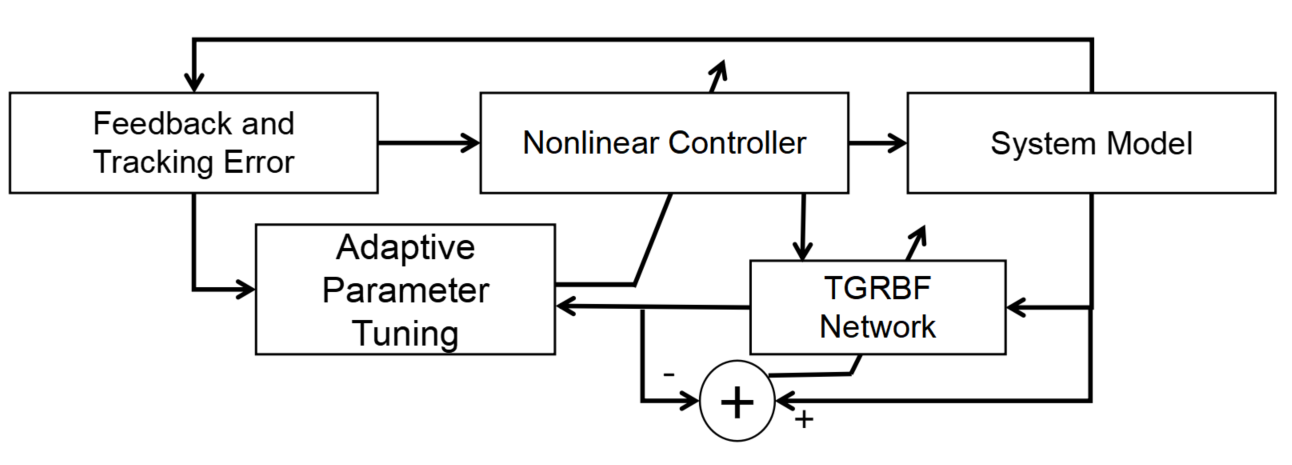}
    \caption{Structure of adaptive control strategy}
    \label{fig3}
\end{figure}

 According to the above control  methods, we can get a complete control strategy, and its process is shown in Fig \ref{fig3}
 
\section{STABILITY ANALYSIS}

In the following, we will demonstrate the stability of the TGRBF-NC  method from three aspects: offline training, online optimization, and control.

\subsection{Convergence  analysis of offline training   }

Firstly , define the Lyapunov function: 
\begin{equation}
    V(t)=\frac{1}{2}e^2(t)
\end{equation}

Where,\(e(t)=f(t)-\hat{f}(t)\),\(\hat{f}(t)\) is is the TGRBF  network fitting system model .

To find the gradient of the Lyapunov function, we get:
\begin{equation}
    \Delta V(t) = V(t+1) - V(t) = \frac{1}{2} \left( e^2(t+1) - e^2(t) \right)
    \label{5}
\end{equation}

The difference of error \(e(t)\) is obtained: 
\begin{equation}
    \Delta e(t)=e(t+1)-e(t)
    \label{2}
\end{equation}

According to the update rules of the network, we can get: 
 \begin{equation}
     e(t+1) = e(t) + \Delta e(t) = e(t) + \left[ \frac{\partial e(t)}{\partial W(t)} \right]^T \Delta W(t) + O(t)
     \label{3}
 \end{equation}
 Where \(W\) is the vector of all the variable parameters in the network, and \(O(t)\) is the higher-order remainder, representing the error 

 The Jacobian matrix of the network is \(J(t) = \frac{\partial \hat{f}(t)}{\partial W(t)}\),so we can get:

 \begin{equation}
     \Delta W(t) = -\eta J(t) e(t)
     \label{1}
 \end{equation}
 
 Bring \eqref{1} in the \eqref{2}   and \eqref{3} to get: 
 \begin{equation}
     \Delta e(t) = \frac{\partial e(t)}{\partial W(t)} (-\eta J(t) e(t)) + O(t)
     \label{4}
 \end{equation}
 
 By substituting the obtained \(\Delta e(t)\) by  \eqref{4}  into \eqref{5}, we get: 
 \begin{equation}
     \begin{split}
         \Delta V(t) =& -\eta J(t)^T e(t)^2 \\&+ \frac{1}{2} \left( \eta^2 J(t)^T e(t)^2 + O(t)^2 + 2\eta J(t)^T e(t) O(t) \right)
     \end{split}
 \end{equation}

Further simplifying the merge, we get: 
\begin{equation}
    \begin{split}
        \Delta V(t) = &-\eta J(t)^T e(t)^2 \left( 1 - \frac{1}{2} \eta J(t)^T \right) \\&+ \frac{1}{2} O(t)^2 + O(t) e(t) \left( 1 - \frac{1}{2} \eta J(t)^T \right)
    \end{split}
    \label{6}
\end{equation}

As we can see from \eqref{6}, when \(0 < \eta P(t)^T < 1\) , The difference component of the Lyapunov function is less than zero, which proves the convergence of off-line training. 

\subsection{Stability analysis of online network parameter optimization    }

We define the augmented state vector and the error term: 
\begin{equation}
    \mathbf{Z}_t = \begin{bmatrix} \widetilde{\mathbf{W}}_t \\ \widetilde{\mathbf{W}}_{t-1} \end{bmatrix}, \quad \widetilde{\mathbf{W}}_t = \mathbf{W}_t - \mathbf{W}^*
\end{equation}

Constructing the Lyapunov function: 
\begin{equation}
V(t) = \frac{1}{2} \| \mathbf{Z}_t \|_F^2 + \frac{1}{2} \| e(t) \|^2    
\end{equation}

where, \(\Gamma = \begin{bmatrix} 1 & -\alpha \\ -\alpha & \alpha^2 \end{bmatrix}\) is a positive definite weight matrix. 

Expanding the difference term :
\begin{equation}
    \begin{split}
        \Delta V(t) &= V(t+1) - V(t)\\& = \frac{1}{2} \left( \| \mathbf{Z}_{t+1} \|_F^2 - \| \mathbf{Z}_t \|_F^2 \right) + \frac{1}{2} \left( \| e(t+1) \|^2 - \| e(t) \|^2 \right)
    \end{split}
    \label{11}
\end{equation}

The expansion of the difference term of the augmented part 
\begin{equation}
    \begin{split}
        &\| \mathbf{Z}_{t+1} \|_F^2 - \| \mathbf{Z}_t \|_F^2 = ( \widetilde{\mathbf{W}}_{t+1}^T \widetilde{\mathbf{W}}_{t+1} - 2\alpha \widetilde{\mathbf{W}}_{t+1}^T \widetilde{\mathbf{W}}_t\\&+ \alpha^2 \widetilde{\mathbf{W}}_t^T \widetilde{\mathbf{W}}_t ) - ( \widetilde{\mathbf{W}}_t^T \widetilde{\mathbf{W}}_t - 2\alpha \widetilde{\mathbf{W}}_t^T \widetilde{\mathbf{W}}_{t-1} + \alpha^2 \widetilde{\mathbf{W}}_{t-1}^T \widetilde{\mathbf{W}}_{t-1} )
    \end{split}
    \label{7}
\end{equation}

By update rule \(W_{t+1} = W_t - \eta_t \nabla J(W_t) + \alpha (W_t - W_{t-1})\),we,can get:
\begin{equation}
    \widetilde{W}_{t+1} = \widetilde{W}_t - \eta_t \nabla J(W_t) + \alpha (\widetilde{W}_t - \widetilde{W}_{t-1})
    \label{8}
\end{equation}

Bring \eqref{8} to \eqref{7} expand and merge to get: 

\begin{equation}
\begin{split}
 &\| \mathbf{Z}_{t+1} \|_F^2 - \| \mathbf{Z}_t \|_F^2  = [ (1+\alpha)^2 \widetilde{\mathbf{W}}_t^T \widetilde{\mathbf{W}}_t + \alpha^2 \widetilde{\mathbf{W}}_{t-1}^T \widetilde{\mathbf{W}}_{t-1} \\&+ \eta_t^2 \| \nabla J \|^2   - 2\alpha (1+\alpha) \widetilde{\mathbf{W}}_t^T \widetilde{\mathbf{W}}_{t-1} - 2(1+\alpha) \eta_t \widetilde{\mathbf{W}}_t^T \nabla J \\&+ 2\alpha \eta_t \widetilde{\mathbf{W}}_{t-1}^T \nabla J   + [ -2\alpha (1+\alpha) \widetilde{\mathbf{W}}_t^T \widetilde{\mathbf{W}}_t + 2\alpha^2 \widetilde{\mathbf{W}}_{t-1}^T \widetilde{\mathbf{W}}_t \\&+ 2\alpha \eta_t \nabla J^T \widetilde{\mathbf{W}}_t ] + \alpha^2 \widetilde{\mathbf{W}}_t^T \widetilde{\mathbf{W}}_t - [ \widetilde{\mathbf{W}}_t^T \widetilde{\mathbf{W}}_t - 2\alpha \widetilde{\mathbf{W}}_t^T \widetilde{\mathbf{W}}_{t-1}\\& + \alpha^2 \widetilde{\mathbf{W}}_{t-1}^T \widetilde{\mathbf{W}}_{t-1} ]
\end{split}
\end{equation}

By simplifying the quadratic, cross and momentum gradient terms and higher order terms, we end up with: 
\begin{equation}
    \begin{split}
      &  \| \mathbf{Z}_{t+1} \|_F^2 - \| \mathbf{Z}_t \|_F^2 = -2 \eta_t \widetilde{\mathbf{W}}_t^T \nabla J + \eta_t^2 \| \nabla J \|^2 \\&+ 2 \alpha \eta_t \left( \widetilde{\mathbf{W}}_{t-1}^T \nabla J + \nabla J^T \widetilde{\mathbf{W}}_t \right)
    \end{split}
\end{equation}

Then, we processed the error difference term. The error term is expanded using Taylor's formula: 
\begin{equation}
    e(t+1) = e(t) + \nabla_e J(\mathbf{W}_t)^T \Delta \mathbf{W}_t + O(\| \Delta \mathbf{W}_t \|^2)
\end{equation}

Further we can get: 
\begin{equation}
    \begin{split}
        &\left\| e(t+1) \right\|^2 - \left\| e(t) \right\|^2 = -2\eta_t e(t)^\top \nabla_e J^\top \nabla J \\&+ 2\alpha e(t)^\top \nabla_e J^\top (\widetilde{\mathbf{W}}_t - \widetilde{\mathbf{W}}_{t-1}) + \eta_t^2 \|\nabla_e J^\top \nabla J\|^2 + O(\eta_t^3)
    \end{split}
\end{equation}

Using the Young inequality, transform the cross term \(2\alpha e(t)^\top \nabla_e J^\top (\widetilde{\mathbf{W}}_t - \widetilde{\mathbf{W}}_{t-1})\).

\begin{equation}
    \begin{split}
        2\alpha e(t)^\top \nabla_e J^\top (\widetilde{\mathbf{W}}_t - \widetilde{\mathbf{W}}_{t-1}) \leq &\frac{\alpha}{\epsilon} \|\nabla_e J^\top (\widetilde{\mathbf{W}}_t - \widetilde{\mathbf{W}}_{t-1})\|^2\\&+\alpha \epsilon \|e(t)\|^2 
    \end{split}
    \label{10}
\end{equation}

We agree that the gradient is satisfied \(\|\nabla_e J\| \leq L\) ,\(L\) is a constant ,then:
\begin{equation}
    \left\| \nabla_e J^\top (\widetilde{\mathbf{W}}_t - \widetilde{\mathbf{W}}_{t-1}) \right\| \leq L \|\widetilde{\mathbf{W}}_t - \widetilde{\mathbf{W}}_{t-1}\|
    \label{9}
\end{equation}

Bring \eqref{9} to \eqref{10},we can get:
\begin{equation}
    \frac{\alpha}{\epsilon} \|\nabla_e J^\top (\widetilde{\mathbf{W}}_t - \widetilde{\mathbf{W}}_{t-1})\|^2 \leq \frac{\alpha L^2}{\epsilon} \|\widetilde{\mathbf{W}}_t - \widetilde{\mathbf{W}}_{t-1}\|^2
\end{equation}

After processing each part of the Lyapunov difference function, we get the total difference term by bringing the processing result into the original formula \eqref{11}
\begin{equation}\begin{aligned}
 \Delta V(t) =& \frac{1}{2} \biggl( -2\eta_{t} \widetilde{\mathbf{W}}_{t}^{T} \nabla J + \eta_{t}^2 \|\nabla J\|^2 \\
& \quad + 2\alpha\eta_{t} \bigl( \widetilde{\mathbf{W}}_{t-1}^{T} \nabla J + \nabla J^{T} \widetilde{\mathbf{W}}_{t} \bigr) \biggr) \\
& + \frac{1}{2} \biggl( -2\eta_{t} e(t)^{T} \nabla_{e}J^{T} \nabla J + \alpha\epsilon \Vert e(t) \Vert^2 \\
& \quad + \frac{\alpha L^2}{\epsilon} \Vert \widetilde{\mathbf{W}}_{t} - \widetilde{\mathbf{W}}_{t-1} \Vert^2 + \eta_{t}^2 \Vert \nabla_{e}J^{T} \nabla J \Vert^2 \biggr) \\
& + O(\eta_{t}^3)
\end{aligned}\label{12}\end{equation}

Due to the explicit step  formula: \(\eta_t = \frac{\mathbf{v}_t^\top \mathbf{F}(\mathbf{W}_t)}{\mathbf{v}_t^\top \mathbf{v}_t}\),we can get:
\begin{equation}
    \eta_t = \frac{\mathbf{F}(\mathbf{W}_t)^\top \nabla_{\mathbf{w}} \mathbf{F}(\mathbf{W}_t) \nabla_{\mathbf{w}} \mathbf{F}(\mathbf{W}_t)^\top \mathbf{F}(\mathbf{W}_t)}{\| \nabla_{\mathbf{w}} \mathbf{F}(\mathbf{W}_t) \nabla_{\mathbf{w}} \mathbf{F}(\mathbf{W}_t)^\top \mathbf{F}(\mathbf{W}_t) \|^2}
\end{equation}
\begin{equation}
      \eta_t\geq \frac{\lambda_{\min}^2}{\lambda_{\max}^4}
\end{equation}

Where \(\lambda\) is the singular value of the Jacobian matrix. We can get :
\begin{equation}
    \| \nabla_{\mathbf{w}} J(\mathbf{W}_t) \|^2 \geq \lambda_{\min^2} \|\widetilde{\mathbf{W}}_t\|^2
\end{equation}
\begin{equation}
     \quad \| \nabla_{\mathbf{e}} J(\mathbf{W}_t) \|^2 \leq \lambda_{\max^2} \|\mathbf{e}(t)\|^2
\end{equation}

Select \(\epsilon = \frac{\lambda_{\min}^2}{2\alpha}\),ignore the higher order terms, substitute \eqref{12} and scale, we can get: 
\begin{equation}
    \begin{split}
        \Delta V(t) \leq &-\eta_t \lambda_{\min^2} \|\widetilde{\mathbf{W}}_t\|^2 + \eta_t^2 \lambda_{\max^2} \|\widetilde{\mathbf{W}}_t\|^2 \\&+ \frac{\alpha L^2 \eta_t}{\epsilon} \|\widetilde{\mathbf{W}}_t - \widetilde{\mathbf{W}}_{t-1}\|^2
    \end{split}
\end{equation}

Again simplify : 
\begin{equation}
    \Delta V(t) \leq -\eta_t \left( \lambda_{\min^2} - \frac{2\alpha^2 L^2}{\lambda_{\min^2}} \right) \| \mathbf{Z}_t \|^2 + \eta_t^2 \lambda_{\max^2} \| \mathbf{Z}_t \|^2
\end{equation}

To guarantee \(\Delta V(t)\leq0\), select 
\begin{equation}
\eta_t \leq \frac{\lambda_{\min^2} - \frac{2\alpha^2 L^2}{\lambda_{\min^2}}}{\lambda_{\max^2}}, \quad \alpha < \frac{\lambda_{\min^2}}{\sqrt{2}L}
\end{equation}

then,existence convergence constant \(\kappa = \frac{\lambda_{\min^4} - 2\alpha^2 L^2}{\lambda_{\max^4}} \in (0,1)\) , let:
\begin{equation}
    \Delta V(t) \leq -\kappa' \eta_t \| \mathbf{Z}_t \|^2
\end{equation}

Combine event trigger conditions \(\|e(t)\|\geq\delta\) ,it is obtained that the finite-time  \(T \leq \frac{V(0)}{\kappa' \delta^2}\) ,network fit is convergent and stable .

\subsection{Stability analysis of adaptive control   }

We first assume that the error between the network fitting system model and the actual system model is \(\epsilon(t) = g(y, u) - \hat{g}(y, u)\) ,
The error has a boundary satisfying  \(\|\epsilon(t)\leq \epsilon_{max}\)

Control tracking reference signal  \(r (t)\), tracking error defined as :  \(e (t) = R(t)-y(t)\). 

Substituting the controller \ref{13} and using the TGRBF network approximation , the error dynamics become: 

\begin{equation}
    e(t+1) = r(t+1) - y(t+1) = r(t+1) - [\hat{g}(\xi(t)) + \epsilon(t) + d(t+1)]
\end{equation}

where  \(\epsilon(t) = g(\xi(t)) - \hat{g}(\xi(t))\) is the approximation error bounded by  \(|\epsilon(t)| \leq \epsilon_{\max}\) .

Rearranging: 

\begin{equation}
    e(t+1) = e(t) + \Delta e(t) + \Delta_d(t+1)
    \label{n1}
\end{equation}
where \(\Delta e(t) = -\frac{\partial \hat{g}}{\partial u} \left( k_1(t)e(t) + k_2(t)\text{sig}^\alpha(e(t)) \right)\) , and \(\Delta_d(t+1) = -\epsilon(t) - d(t+1)\) , bounded by \(|\Delta_d(t)| \leq \epsilon_{\max} + \bar{D}\) .

Design the Lyapunov function: 
\begin{equation}
    V(t) = \frac{1}{2} e^2(t) + \frac{1}{2\eta_1} \tilde{k}_1^2(t) + \frac{1}{2\eta_2} \tilde{k}_2^2(t)
\end{equation}

where  \(\tilde{k}_i = k_i - k_i^*\) are parameter errors, and \(k_i^*\) are ideal gains. The Lyapunov difference is :

\begin{equation}
    \Delta V(t) = V(t+1) - V(t)
\end{equation}

Expanding each term: 

\begin{equation}
\begin{split}
     \Delta V(t) = &\frac{1}{2} \left( e^2(t+1) - e^2(t) \right) + \frac{1}{2\eta_1} \left( \tilde{k}_1^2(t+1) - \tilde{k}_1^2(t) \right)\\& + \frac{1}{2\eta_2} \left( \tilde{k}_2^2(t+1) - \tilde{k}_2^2(t) \right)
\end{split}
\end{equation}

Substitute \ref{n1} :

\begin{equation}
\begin{split}
      \frac{1}{2} \left( e^2(t+1) - e^2(t) \right) =& e(t)\Delta e(t) + \frac{1}{2} \Delta e^2(t) \\&+ e(t)\Delta_d(t+1) + \frac{1}{2} \Delta_d^2(t+1) 
\end{split}
\end{equation}

Using the  control gain adaptive laws \ref{n2} and \ref{n3} , the parameter errors evolve as: 
\begin{equation}
    \Delta \tilde{k}_i(t) = \tilde{k}_i(t+1) - \tilde{k}_i(t) = \eta_i \frac{\partial y_m}{\partial u} 
\begin{cases} 
e^2(t) & \text{for } i = 1 \\
e(t)\text{sig}^\alpha(e(t)) & \text{for } i = 2 
\end{cases}
\end{equation}
Substitute into the parameter error terms :

\begin{equation}
    \frac{1}{2\eta_i} \left( \tilde{k}_i^2(t+1) - \tilde{k}_i^2(t) \right) = \tilde{k}_i(t) \Delta \tilde{k}_i(t) + \frac{1}{2\eta_i} \Delta \tilde{k}_i^2(t)
\end{equation}

Combine all terms: 

\begin{equation}
\begin{split}
    \Delta V(t) =& e(t)\Delta e(t) + \frac{1}{2}\Delta e^2(t) + e(t)\Delta_d(t+1) +\\& \frac{1}{2}\Delta_d^2(t+1) + \sum_{i=1}^{2} \left( \tilde{k}_i(t)\Delta \tilde{k}_i(t) + \frac{1}{2\eta_i}\Delta \tilde{k}_i^2(t) \right)  
\end{split}
\end{equation}

From \(\Delta e(t) = -L_u \left( k_1(t)e(t) + k_2(t)|e(t)|^\alpha \right)\) , where \(L_u = \left| \frac{\partial \hat{g}}{\partial u} \right|\) :

\begin{equation}
    e(t)\Delta e(t) = -L_u k_1(t)e^2(t) - L_u k_2(t)|e(t)|^{\alpha+1}
\end{equation}

Using \(|\Delta e(t)| \leq L_u \left( k_1(t)|e(t)| + k_2(t)|e(t)|^\alpha \right)\) :

\begin{equation}
\begin{split}
       \frac{1}{2} \Delta e^2(t) \leq &\frac{L_u^2}{2} ( k_1^2(t)e^2(t) + 2k_1(t)k_2(t)|e(t)|^{\alpha+1} \\&+ k_2^2(t)|e(t)|^{2\alpha} )
\end{split}
\end{equation}

Apply Young's inequality :

 \begin{equation}
     e(t)\Delta_d(t+1) \leq \frac{1}{2}e^2(t) + \frac{1}{2}(\epsilon_{\max} + \bar{D})^2
 \end{equation}

 Using  \(\Delta \tilde{k}_i^2(t) \leq \eta_i^2 \left| \frac{\partial y_m}{\partial u} \right|^2 
\begin{cases} 
e^4(t) & \text{for } i = 1 \\
e^2(t)|e(t)|^{2\alpha} & \text{for } i = 2 
\end{cases}\) :

\begin{equation}
    \frac{1}{2\eta_i} \Delta \tilde{k}_i^2(t) \leq \frac{\eta_i}{2} \left| \frac{\partial y_m}{\partial u} \right|^2 
\begin{cases} 
e^4(t) & \text{for } i = 1 \\
e^2(t)|e(t)|^{2\alpha} & \text{for } i = 2 
\end{cases}
\end{equation}

Substitute all bounds into \(\Delta V(t)\) :

\begin{equation}
\begin{split}
        \Delta V(t) \leq &-L_u k_1(t)e^2(t) - L_u k_2(t)|e(t)|^{\alpha+1} \\
        & + \frac{L_u^2}{2}k_1^2(t)e^2(t)+ L_u^2 k_1(t)k_2(t)|e(t)|^{\alpha+1} \\
        &+ \frac{L_u^2}{2} k_2^2(t)|e(t)|^{2\alpha}  + \frac{1}{2}e^2(t) + \frac{1}{2}(\epsilon_{\max} + \bar{D})^2  \\
        &+ \sum_{i=1}^{2} ( \eta_i \frac{\partial y_m}{\partial u} \tilde{k}_i(t) 
\begin{cases} 
e^2(t) & \text{for } i = 1 \\
e(t)|e(t)|^\alpha & \text{for } i = 2 
\end{cases}   \\
        &+ \frac{\eta_i}{2} \left| \frac{\partial y_m}{\partial u} \right|^2 
\begin{cases} 
e^4(t) & \text{for } i = 1 \\
e^2(t)|e(t)|^{2\alpha} & \text{for } i = 2 
\end{cases} )
\end{split}
\end{equation}

To ensure \(\Delta V(t) \leq -\kappa V(t) + C\) ,group terms and enforce negativity. we can choose :

\begin{align*}
k_1(t) &> \frac{1 + \sqrt{1 + 2L_u^2}}{2L_u}, \\
k_2(t) &> \frac{L_u k_1(t)}{1 - L_u k_1(t)}, \\
\eta_1 &< \frac{2}{|\frac{\partial y_m}{\partial u}|^2}, \\
\eta_2 &< \frac{2}{|\frac{\partial y_m}{\partial u}|^2}.
\end{align*}

The constant \(C\) aggregates disturbances :

\begin{equation}
    C = \frac{1}{2}(\epsilon_{\max} + \bar{D})^2 + \sum_{i=1}^{2} \frac{\eta_i^2}{2}
\end{equation}

By the discrete-time Lyapunov theorem, if the above conditions hold, the tracking error \(e(t)\)  and parameter errors \(\tilde{k}_i(t)\) are uniformly ultimately bounded: 

\begin{equation}
    \limsup_{t \to \infty} |e(t)| \leq \sqrt{\frac{2C}{\kappa}}
\end{equation}

where 

\begin{equation}
    \kappa = \min ( 2L_u k_1(t) - L_u^2 k_1^2(t) - 1, \, 2L_u k_2(t) - 2L_u^2 k_1(t)k_2(t) )
\end{equation}

 In further discussion, the fast Lyapunov function is constructed when \(C=0\), that is, the fitting of the network is accurate enough 

 \begin{equation}
     V(t) = \frac{1}{2} e^2 + \frac{1}{\gamma_1 (\alpha + 1)} |\tilde{k}_1|^{\alpha+1} + \frac{1}{\gamma_2 (\alpha + 1)} |\tilde{k}_2|^{\alpha+1}
 \end{equation}

 According to lemma II.3 of \cite{b18} , there are limited time \(T \leq \frac{1}{c(1-\alpha)} \ln \left( 1 + \frac{c V^{1-\alpha}(0)}{C} \right)\),make \(e (t) \rightarrow 0\), globally finite time stable system. 
\section{Simulation experiment }

To verify the effectiveness of the proposed adaptive control method based on the TGRBF network, this section conducts multi-dimensional performance tests through simulation experiments and compares the results with those of traditional methods. The experiments are divided into two parts: verification of model fitting accuracy and comparison of control performance. 

\subsection{Controlled system and experimental setup    }
According to the above Settings, we design the following nonlinear system as the controlled system:
\begin{equation}
    \left\{
\begin{aligned}
x_1(k+1) &= x_2(k) \\
x_2(k+1) &= \frac{1}{T} \left( -2x_2(k) - \sin(x_1(k)) + u(k-1) \right) + d(k) \\
y(k) &= K x_1(k)
\end{aligned}
\right.
\end{equation}

In the case of the nonlinear system, the state delay \(n_y=2\) , and the input delay \(n_d=1\). During the offline fitting process, to highlight the differences between the used simulation model and the actual nonlinear system, adjustments were made to the parameters and unmodeled disturbances were added. The actual system parameters are \(T=3, K=0.5\), while the parameters of the ideal model used for fitting are \(T=2, K=2\). In addition to white noise, the actual system \(d(k)\) also has sinusoidal low-frequency disturbances. The system sampling time \(T_s= 0.001s\) , and the initial states are all 0.

The TGRBF network has \(m=6\) hidden nodes, a GRU hidden state dimension of \(p=6\), and an initial gate weight of 0.5. The initial values of the controller gains are \(k1=1.5, k2=0.8\), and \(\alpha=0.7\). The event-triggered threshold for online network optimization is 0.01, and the momentum factor is set to 0.2.

In the model fitting accuracy test, comparisons are made using DNN, LSTM, RBFNN and TGRBF networks. For the control accuracy test, comparisons are conducted using PID, NC, and TGRBF-NC methods. All methods are tested in the same environment.

\subsection{Network performance test }

In the ideal model, following the data acquisition method mentioned above , 1000 sets of input-output data were generated and organized into a training dataset in chronological order. A random continuous time-series input was designed for testing. 

\begin{figure}
    \centering
    \includegraphics[width=1\linewidth]{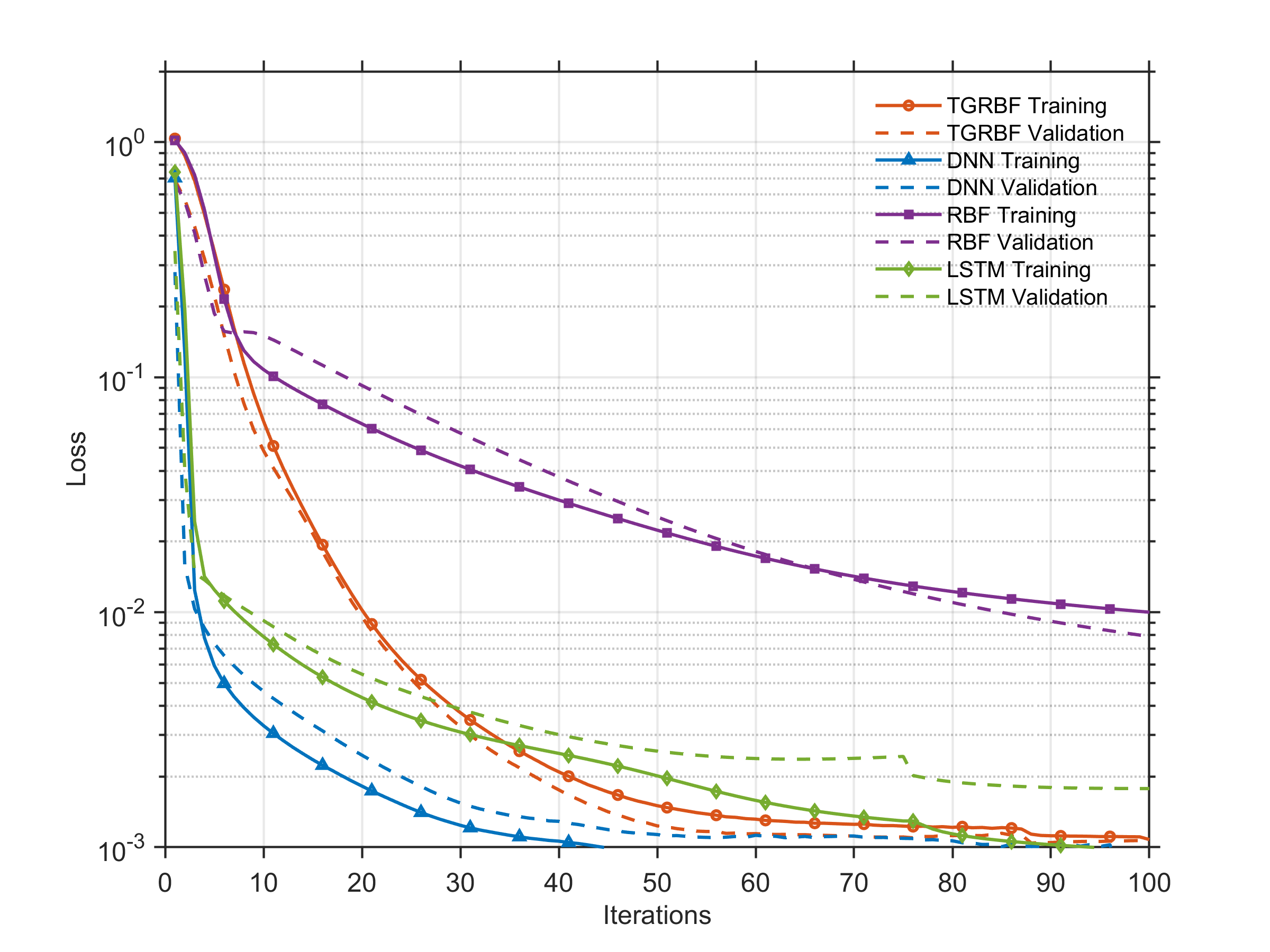}
    \caption{Training loss comparison of different networks during pre-training}
    \label{fig4}
\end{figure}

The Fig \ref{fig4} visualizes the training loss comparison of different networks during pre-training. It can be seen from the figure that the DNN has the fastest loss reduction, but the TGRBF also brought the loss down to the same level as the DNN at the end of the training. The RBF network and the LSTM network still had higher losses at the end of the training, indicating poor fitting.

The Table I compares and analyzes the number of parameters and fitting performance of different networks. MSE,RMSE,MAE and R2 are used for performance quantification to ensure accurate analysis. As can be seen from the table, TGRBF achieves the best results with a relatively small number of parameters. 

\begin{table}
\centering
\caption{ The number of parameters and fitting performance of different networks}
\begin{tabular}{| l | l | l | l | l | l |}
\hline
\textbf{\textbf{Model}} & \textbf{\textbf{Parameters}} & \textbf{\textbf{MSE}} & \textbf{\textbf{RMSE}} & \textbf{\textbf{MAE}} & \textbf{\textbf{R2}} \\
\hline
TGRBF & 332 & 0.0062& 0.0791& 0.0311& 0.9989\\
\hline
DNN & 786 & 0.0078& 0.0885& 0.0621& 0.9985\\
\hline
RBF & 290 & 0.0625& 0.2501& 0.1349& 0.9878
\\
\hline
LSTM & 3906 & 0.0099& 0.0989& 0.0637& 0.9981\\
\hline

\end{tabular}

\end{table}

\begin{figure}
    \centering
    \includegraphics[width=1\linewidth]{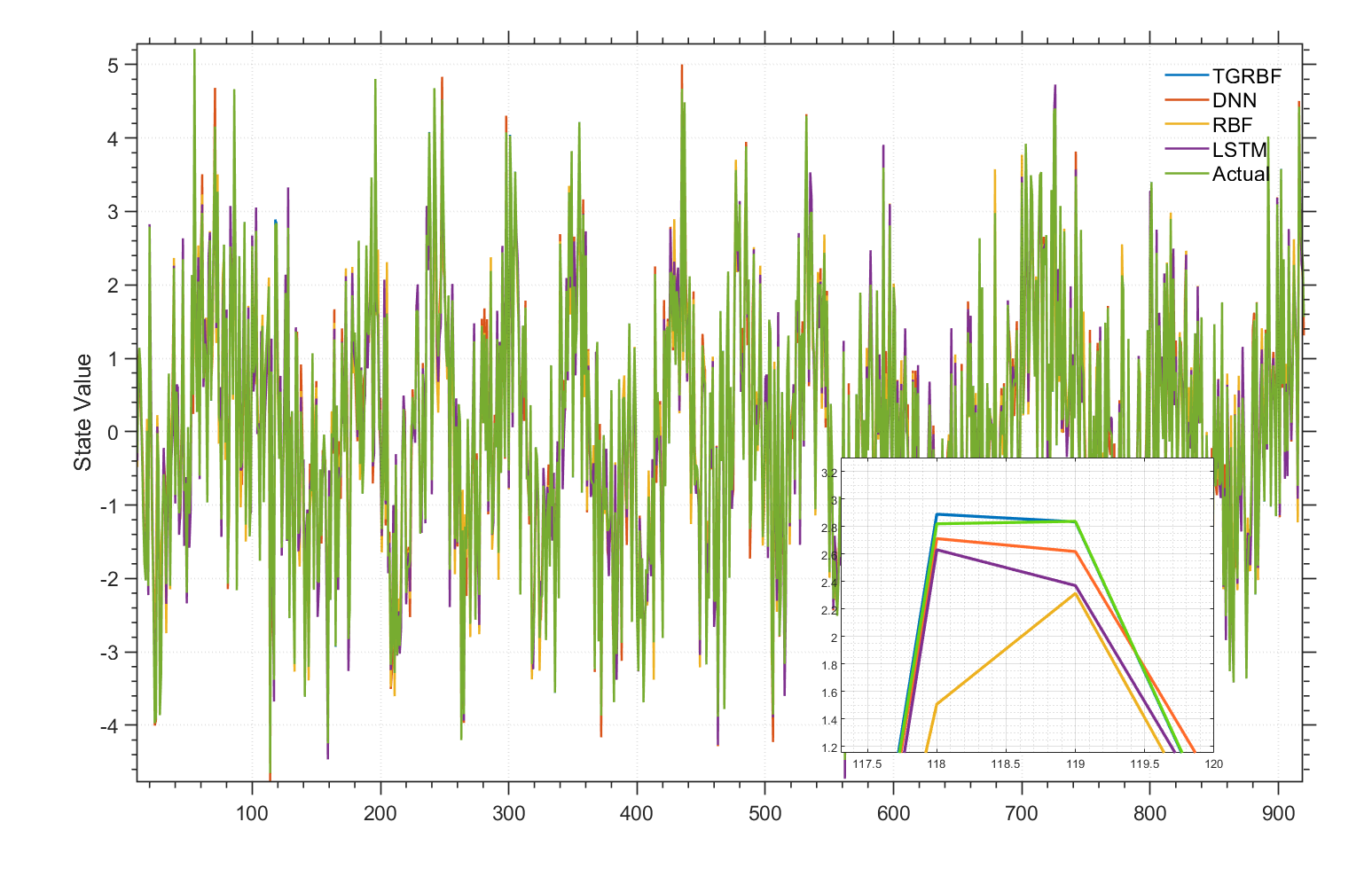}
    \caption{the predicted system output of different networks under the same input sequence with the output of the actual system}
    \label{fig5}
\end{figure}

Fig.\ref{fig5}  compares the predicted system output of different networks under the same input sequence with the output of the actual system. The predicted output of TGRBF is highly consistent with the actual system (MSE=0.006), which verifies its effectiveness in nonlinear dynamic modeling .

\subsection{Control performance test }

Following the system identification tests, comprehensive evaluations of control performance were conducted using step reference signals and sinusoidal tracking trajectories to compare three control methodologies: conventional PID, nonlinear control (NC), and the proposed neural network-enhanced control (NN-NC).

\begin{figure}
    \centering
    \includegraphics[width=1\linewidth]{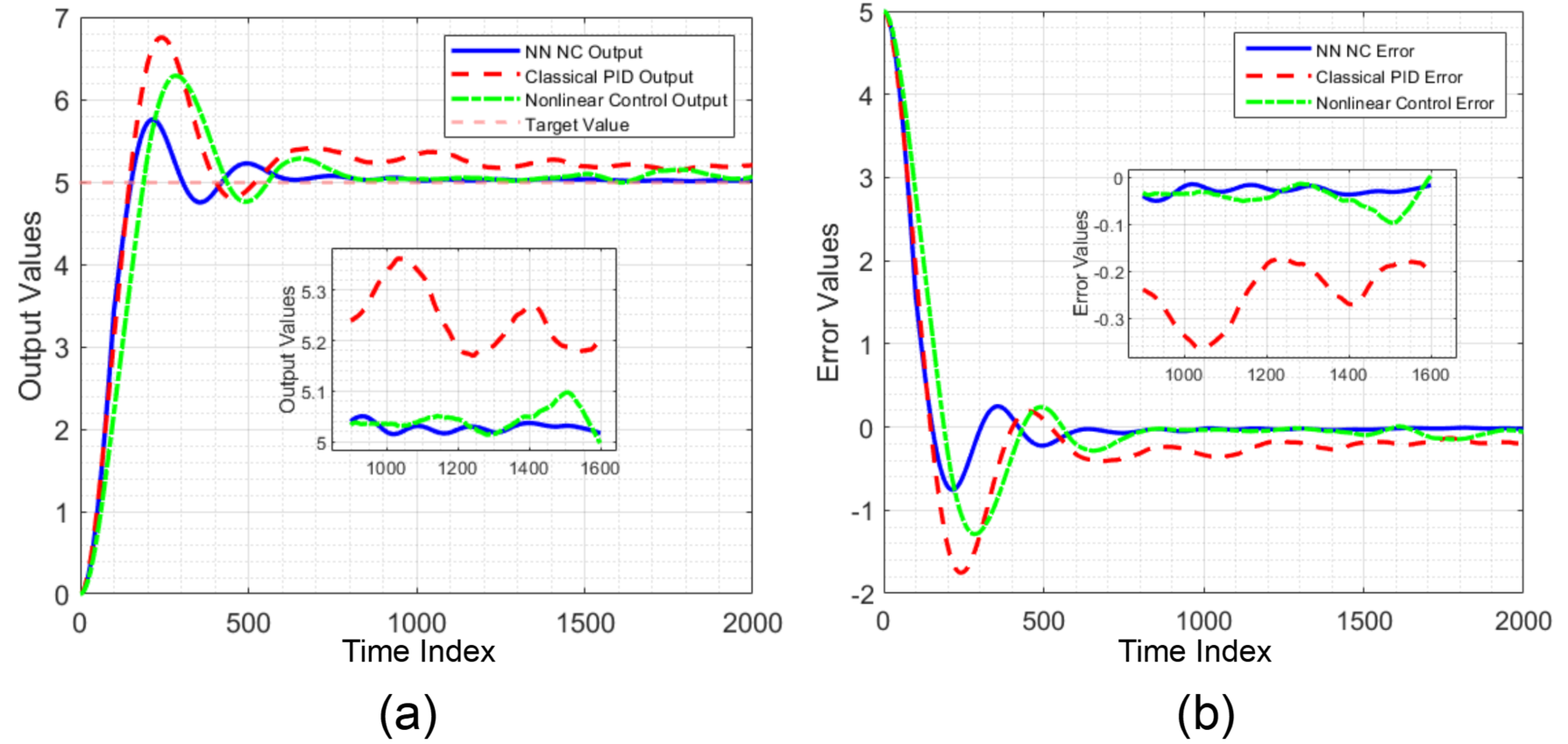}
    \caption{Control performance of different controllers under step reference signal}
    \label{fig6}
\end{figure}

\begin{figure}
    \centering
    \includegraphics[width=1\linewidth]{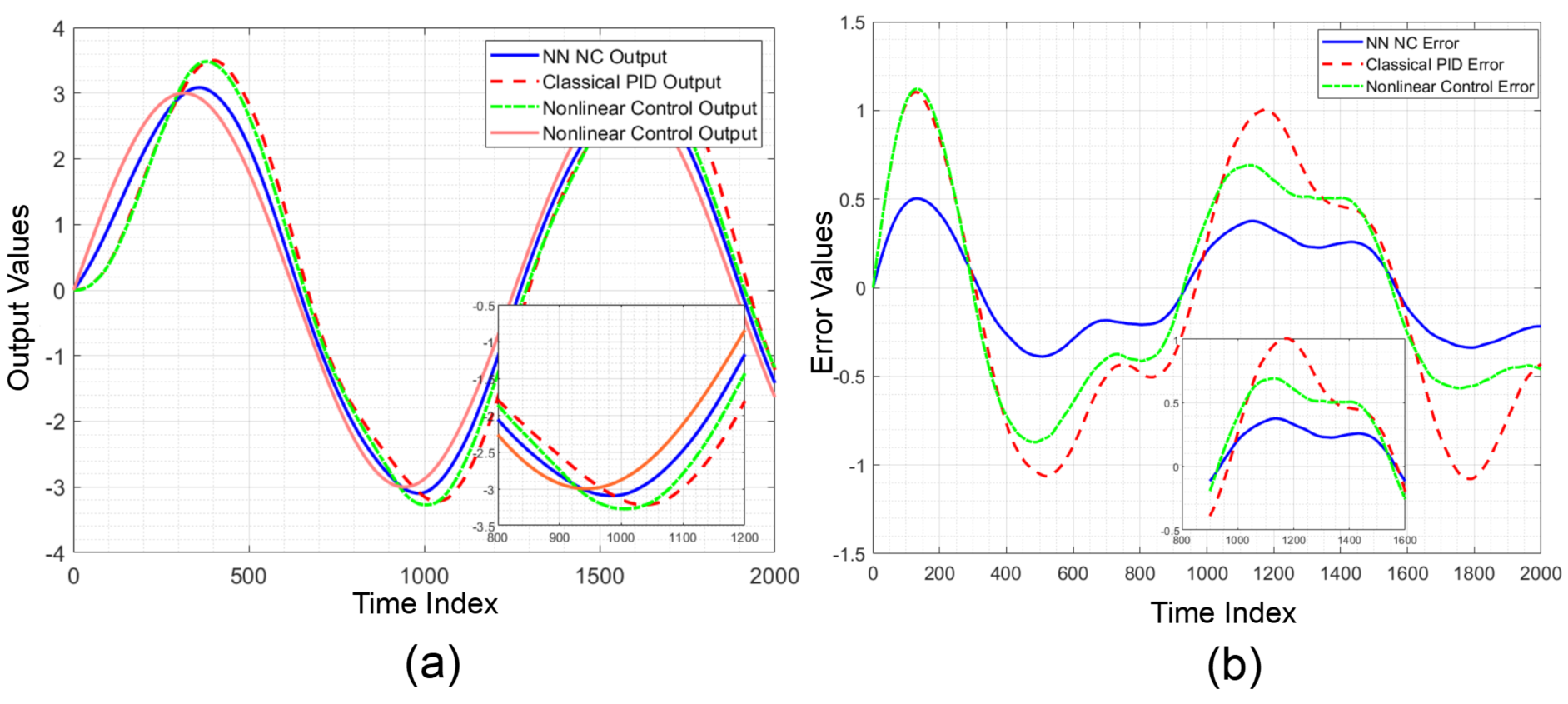}
    \caption{Control performance of different controllers under sinusoidal tracking trajectories}
    \label{fig7}
\end{figure}

As illustrated in the tracking results (Fig. \ref{fig6} , Fig.\ref{fig7} ), the PID controller exhibited significant oscillations and steady-state deviations due to uncompensated nonlinearities and external disturbances. While the NC demonstrated improved robustness and faster transient response compared to PID, its fixed-gain structure resulted in excessive overshoot (25.7\%) and residual tracking errors under time-varying conditions. In contrast, the NN-NC achieved near-optimal tracking with merely 11.1\% overshoot and stabilized within 0.7 seconds, demonstrating superior adaptability to nonlinear dynamics and disturbance rejection.

For quantitative analysis, five performance metrics were computed: Integral Absolute Error (IAE), Integral Squared Error (ISE), Integral Time-weighted Absolute Error (ITAE), maximum overshoot, and settling time (2\% criterion). As summarized in Table II ,Table III . the NN-NC reduced IAE by 28.8\% and ITAE by 47.6\% relative to NC, while achieving 60.6\% faster settling than PID. These results validate the dual advantages of the TGRBF-based adaptation mechanism in simultaneously enhancing transient performance and steady-state precision under dynamic operating conditions.

\begin{table}
\centering
\caption{performance metrics for step reference signal tracking }
\begin{tabular}{| l | l | l | l |}
\hline
\textbf{\textbf{Metric}} & \textbf{\textbf{NN NC}} & \textbf{\textbf{Classical PID}} & \textbf{\textbf{Nonlinear Control}} \\
\hline
IAE & 7.7642& 10.6869& 8.7415
\\
\hline
ISE & 16.3862& 20.8624& 22.6212
\\
\hline
ITAE & 17.0391& 51.7214& 23.5461\\
\hline
Overshoot (\%) & 11.1241& 35.022& 25.7261
\\
\hline
Settling Time (s) & 0.701& 1.781& 0.772\\
\hline

\end{tabular}
\label{tab2}
\end{table}

\begin{table}
\centering
\caption{performance metrics for sinusoidal  tracking}
\begin{tabular}{| l | l | l | l |}
\hline
\textbf{\textbf{Metric}} & \textbf{\textbf{Neural\_Network}} & \textbf{\textbf{Classical\_PID}} & \textbf{\textbf{Nonlinear\_Control}} \\
\hline
IAE & 7.4441& 12.7466& 10.3981\\
\hline
ISE & 3.3876& 9.9173& 6.4962
\\
\hline
ITAE & 46.3199& 125.3709& 94.1457\\
\hline

\end{tabular}
\label{tab3}
\end{table}

\begin{figure}
    \centering
    \includegraphics[width=1\linewidth]{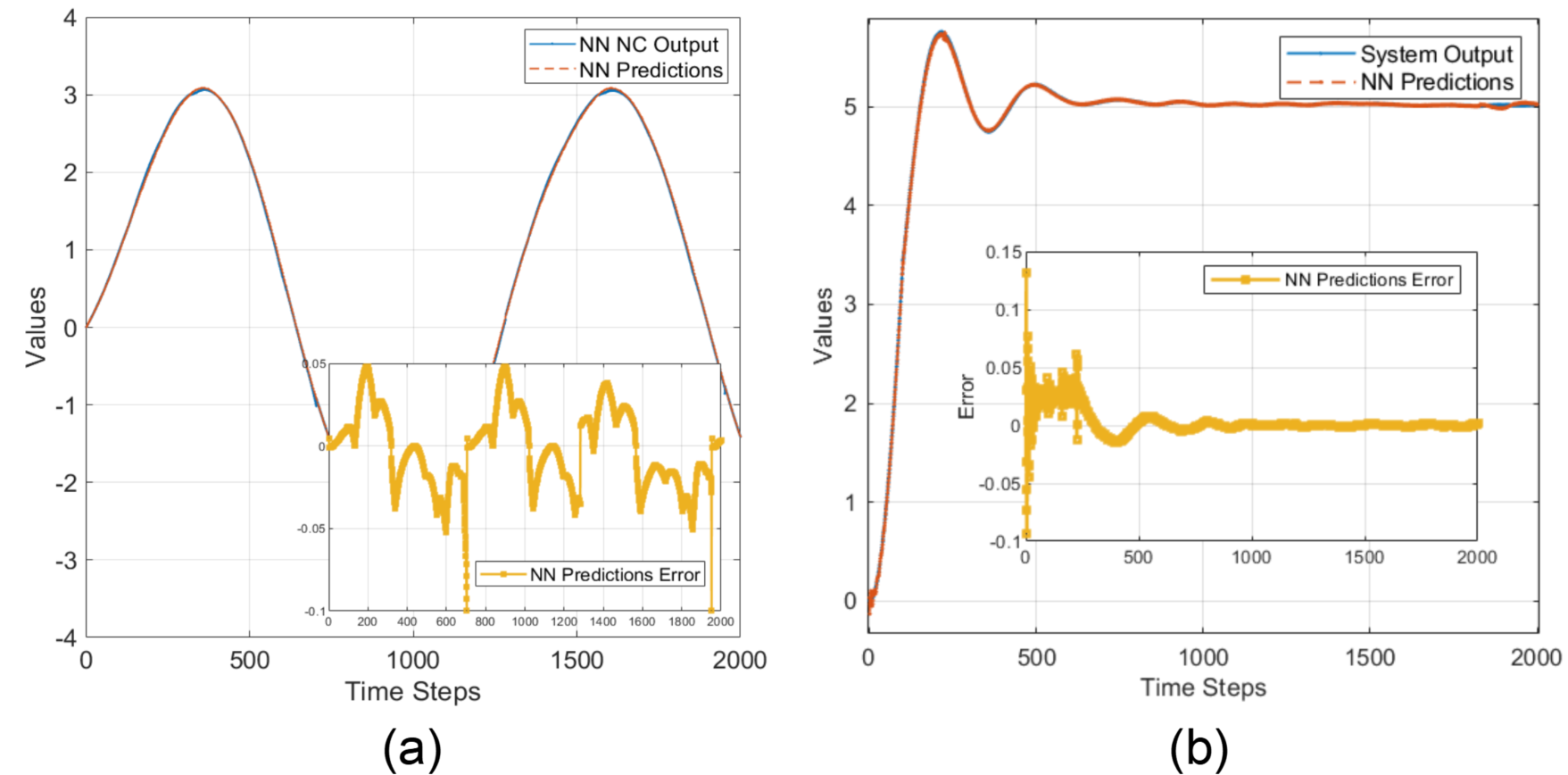}
    \caption{the comparison between the predicted system output and the actual system output of the network in the control process of sinusoidal tracking (a) and step tracking (b)}
    \label{fig9}
\end{figure}

The  Fig.\ref{fig9}  shows the comparison between the predicted system output and the actual system output of the network in the control process of sinusoidal tracking and step tracking, and it can be seen that the network fits the output of the system very accurately in the control process. And when there is a large error, the online optimization is started to quickly eliminate the error to ensure the accuracy of the network. 

\section{Conclusion}
This paper presents a novel adaptive control framework based on a Temporal Gated Radial Basis Function (TGRBF) network, which addresses the limitations of conventional methods in handling unknown nonlinear dynamics and time-varying disturbances by synergizing the local approximation strength of RBF neural networks (RBFNNs) with the temporal modeling capabilities of gated recurrent units (GRUs). Building upon this architecture, we further develop an event-triggered adaptive high-performance control strategy. The key innovations are threefold:  

1. Lightweight Hybrid Architecture: A dynamic gating mechanism enables long-term temporal modeling with only 16.2\% additional parameters, significantly reducing computational complexity compared to conventional approaches.  

2. Event-Triggered Online Optimization: Momentum-explicit gradient descent combined with experience replay sampling achieves optimal trade-offs between real-time responsiveness and system stability.  

3. Formal Stability Guarantees: Lyapunov-based analysis rigorously proves the uniform ultimate boundedness of both tracking errors and network parameters.  

Simulation studies demonstrate that the TGRBF-enhanced control (TGRBF-NC) outperforms deep neural networks (DNNs), LSTM-based controllers, and classical control methods in both modeling accuracy (RMSE = 0.079) and control performance (overshoot = 11.1\%), while maintaining a parameter count at merely 8.5\% of conventional LSTM architectures. Future research will extend the TGRBF framework to robotic trajectory tracking and multi-agent cooperative control scenarios, with additional optimizations targeting real-time learning efficiency for embedded implementations. This work bridges the gap between data-driven adaptability and model-based stability guarantees, offering a principled solution for complex systems operating under dynamic uncertainties.

\bibliographystyle{IEEEtran}
\bibliography{ref}

\end{document}